\documentclass{article}
\usepackage{amsmath}
\usepackage{amsfonts}
\usepackage{amssymb}
\usepackage{graphicx}

\title{Dynamical Correlations as Origin of Nonextensive Entropy}

\author{
  T. Kodama$^{1}$, H.-T. Elze$^{1}$, C. E. Aguiar$^{1}$
  and T. Koide$^{2}$
\\
\\
  $^{1}$ Instituto de F\'{\i}sica, Universidade Federal do Rio de Janeiro\\
  $^{2}$ Institut f\"{u}r Theoretische Physik, University of Frankfurt
}

\date{}

\begin{document}
\maketitle

\begin{abstract}
We present a simple and general argument showing that a class of dynamical
correlations give rise to the so-called Tsallis nonextensive statistics. An
example of a system having such a dynamics is given, exhibiting a
non-Boltzmann energy distribution. A relation with prethermalization
processes is discussed.
\end{abstract}

\section{Generalized statistics and correlations}

In the recent paper by J.~Berges~et~al. \cite{Berges}, the authors show that
the dynamics of a quantum scalar field exhibits a prethermalization
behavior, where the thermodynamical relations become valid long before the
real thermal equilibrium is attained. There, although the system is
spatially homogeneous, the single particle spectra are different from those
determined by the usual Boltzmann equilibrium. This means that the validity
of thermodynamics does not imply thermal equilibrium in the sense of
Boltzmann.

Such a situation has also long been claimed by Tsallis \cite%
{Tsallis,TsallisRev}, who introduced the $q$-entropy, defined in terms of
the occupation probabilities $\left\{ p_{\alpha }\right\}$ of the
microstate $\alpha$ by 
\begin{equation}
S_{q}\left[ \left\{ p_{\alpha }\right\} \right] =\frac{1-\sum_{\alpha
}p_{a}^{q}}{q-1}\, ,  
\label{STsallis}
\end{equation}
instead of the usual entropy 
$S\left[ \left\{ p_{\alpha }\right\} \right] 
=-\sum_{\alpha }p_{\alpha }\ln p_{\alpha }$.
All thermodynamical relations are derived from the maximization of this 
$q$-entropy, 
\begin{equation}
\delta S_{q}-\frac{1}{T}\delta \left\langle E\right\rangle _{q}+\lambda
\delta \sum_{\alpha }p_{\alpha }=0 \, ,
\end{equation}%
where 
\begin{equation}
\left\langle E\right\rangle _{q}=\frac{\sum_{\alpha }p_{\alpha
}^{q}E_{\alpha }}{\sum_{\alpha }p_{\alpha }^{q}}
\end{equation}%
is the $q$-biased average of the energy. It is shown that, although the
thermodynamical Legendre structure is maintained, the energy spectrum for
such a system is not given by Boltzmann statistics but behaves
asymptotically like a power in $E$ \cite{Tsallis,TsallisRev}. In other
words, the thermal equilibrium in the sense of Boltzmann is a sufficient
condition, but not a necessary one for the thermodynamic relations to be
valid. Extensive studies have been developed \cite{TsallisRev} and the
applications of Tsallis' entropy extend to biological systems, high-energy
physics, and cosmology. However, in spite of these studies, the dynamical
origin of the $q$-entropy and $q$-biased average of observables is not clear
yet.

It has been suggested that Tsallis' nonextensive statistics applies to a
kind of metastable state which is realized before the true thermal
equilibrium, or to some stationary state having long-range correlation \cite%
{TsallisRev}. This is exactly another important result of J.~Berges~et~al.,
that their preequilibrium state is attained long before the real equilibrium
is reached, that is, the system seems to attain first this preequilibrium,
then slowly goes to the real equilibrium. This reminds us somewhat of the
following situation which we encounter frequently in optimization problems.
For example, let us consider the problem of finding a minimum of a function
of two variables, as shown in Fig. 1.

\begin{figure}
\begin{center}
\includegraphics[width=7.0cm]{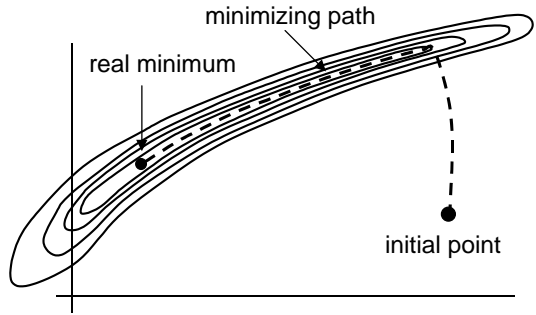}
\caption{Schematic illustration of the effect of a strong correlation among
two variables for minimizing a function.}
\end{center}
\end{figure}

When the function has a narrow long valley, the usual steepest descent
method always first goes to the closest point of the valley instead of going
directly to the real minimum, as indicated by the dashed curve. The narrow
valley means that there exists a strong correlation between the two
variables. In other words, if there exist any strong correlations among
variables, the minimizing path always tries to satisfy the correlation
first, then follows downstream to the true minimum along the valley
generated by the correlation. If the correlation is strong enough, non
Boltzmann distribution may result as the stable configuration of the system
(see Ref. \cite{EK}).

The above consideration suggests that a preequilibrium state may correspond
to a kind of narrow valley of the free energy of some subsystem generated by
dynamical correlations of the system. If such a local minimum stays for a
very long time, we may consider the system as in equilibrium, but the
microscopic occupation probability is not that of the Maxwell-Boltzmann
distribution. Then what characterizes such a local equilibrium? We suggest
here that a general form of Tsallis' statistics is one answer. Of course,
prethermalization involves a broad range of phenomena, and has been studied
for a long time in several works (for recent references, see \cite{NonEquil}%
). Here we concentrate on the discussion of how the states described by the
Tsallis nonextensive statistics can be regarded as a kind of prethermalized
states.

Let us consider a system composed of $N$ particles and let $\left\{
E_{\alpha }\right\} $ denote the single-particle spectrum of the system and $%
\left\{ N_{\alpha }\right\} $ be the corresponding occupation numbers. We
further assume that the system is dilute enough that the interaction
energies are negligible. Then, by definition, 
$E =\sum_{\alpha }N_{\alpha }E_{\alpha }$ 
and 
$N =\sum_{\alpha }N_{\alpha }$
are, respectively, the total energy and total number of particles of the
system. Suppose there exists a strong correlation among any $q$ particles of 
$N_{\alpha }$ in any single-particle state $\alpha$. Then the number of
possible ways of forming such correlated subsystems in each $\alpha $ state
is given by 
\begin{equation}
M_{\alpha }=\left( 
\begin{array}{c}
N_{\alpha } \\ 
q%
\end{array}%
\right) \approx \frac{N_{\alpha }^{q}}{\Gamma (q+1)}  \label{M_alfa}
\end{equation}%
for $N_{\alpha }\gg q$. Note that here, $q$ does not have to be integer. We
call these correlated systems $q$-clusters. The total number of ways of
forming $q$-cluster states for the whole system is 
\begin{equation}
M=\sum_{\alpha }M_{\alpha } \,.
\end{equation}

If the $q$-clusters are formed with equal a priori probability, thus
proportional to the number of ways of forming them, then the average energy
per cluster particle is 
\begin{equation}
\left\langle E\right\rangle _{q}=
\frac{\sum_{\alpha }M_{\alpha }E_{\alpha }}{\sum_{\alpha }M_{\alpha }}=
\frac{\sum_{\alpha }p_{\alpha }^{q}E_{\alpha }}{\sum_{\alpha }p_{\alpha}^{q}},  
\label{E_q}
\end{equation}%
where $p_{\alpha }=N_{\alpha }/N$ is the probability of a particle occupying
state $\alpha .$

Now, we further assume that the dynamical evolution of the system is such
that, for a given number of $q$-clusters, the configuration of the system
tries to minimize the energy of the correlated subsystem. That is, for some
arbitrary initial condition, the system first generates correlations among
particles in such a way to minimize the energy in the clusters. The narrow
valley in Fig.~1 corresponds to these ``ground states'' of the subsystem
formed by $q$-clusters. Neglecting the component perpendicular to the narrow
valley, the equilibrium corresponds to minima of the expectation value $%
\left\langle E\right\rangle _{q}$, 
\begin{equation}
\delta \left\langle E\right\rangle _{q}=0 \, ,  \label{dE=0}
\end{equation}%
under the conditions 
\begin{align}
\sum_{\alpha }N_{\alpha }& =N \, ,  \label{N} \\
\sum_{\alpha }p_{\alpha }^{q}& =\frac{M}{N^{q}}=A=const.  \label{A}
\end{align}
This is equivalent to the variational problem, 
\begin{equation}
\delta \left[ \frac{\sum_\alpha p_\alpha^q E_\alpha}{\sum_{\alpha}p_{\alpha
}^q}\right. + \lambda \left( \sum_{\alpha }p_{\alpha }^{q}-A\right) 
+ \left. \nu \left( \sum_{\alpha}p_{\alpha }-1\right) \right] =0 \, ,
\label{min}
\end{equation}%
which determines the probabilities $p_{\alpha }$. We obtain: 
\begin{equation}
p_{\alpha }=\frac{1}{Z_{q}}\left[ 1+\frac{1}{\lambda A}(E_{\alpha
}-\left\langle E\right\rangle _{q})\right] ^{1/(1-q)},  \label{p_a}
\end{equation}%
where $Z_{q}=A^{1/(1-q)}$ is a normalization factor.
We see that this expression is nothing but the Tsallis distribution
function, if we put $\lambda ^{-1}=(q-1)\beta $. Note that there was no need
to introduce the concept of entropy yet. The Tsallis entropy here
corresponds to the simple normalization condition for the number of $q$%
-clusters.

It is important to note that the minimization condition, Eq.(\ref{min}), of
the energy of the correlated subsystems under the constraint Eq.(\ref{A}) 
determines completely the occupation probabilities for the whole system as
functions of $A$ and of the single-particle spectrum, 
$p_{\alpha }=p_{\alpha }\left( A,\left\{ E_{i}\right\} \right)$.
When we change the external parameters of the system, such as the volume 
$V$, then the single-particle spectrum will suffer a change. However, if the
dynamics of the system is such that the number of correlated particles $q$
is kept constant during slow changes of $V$ and $N$, then we can consider 
$V$, $N$, and $A$ independent ``state''
parameters, since the occupation probabilities for the system are completely
determined by these variables. We can thus calculate the total energy as
function of them as 
\begin{equation}
E(V,A,N)=N\sum_{\alpha }p_{\alpha }E_{\alpha } \, .
\end{equation}%
Comparing the variation of this energy function with the First Law of
Thermodynamics, $\delta E=-P\delta V+T\delta S+\mu \delta N$,
we can identify the pressure 
\begin{equation}
P=-N\sum_{\alpha }\left( \sum_{\beta }\left( \frac{\partial p_{\beta }}{%
\partial E_{\alpha }}\right) _{A}E_{\beta }+p_{\alpha }\right) \frac{%
\partial E_{\alpha }}{\partial V} \, ,  \label{P}
\end{equation}%
the chemical potential  
$\mu =\sum_{\alpha }p_{\alpha }E_{\alpha }$, 
and the entropy term 
\begin{equation}
T\delta S=\left[ N\sum_{\alpha }\left( \frac{\partial p_{\alpha }}{\partial A%
}\right) _{V}E_{\alpha }\right] \delta A \, .  \label{Tds=FdA}
\end{equation}%
Eq.~(\ref{Tds=FdA}), implies that we can relate the parameter $A$ to the
entropy of the system, up to an arbitrary function $f(A)$, 
\begin{equation}
S=f(A) \, ,
\end{equation}%
and have a temperature given by 
\begin{equation}
T=\frac{1}{f^{\prime }\left( A\right) }\left[ \sum_{\alpha }\left( \frac{%
\partial p_{\alpha }}{\partial A}\right) _{V}E_{\alpha }\right] \, .
\end{equation}%
If we take $f(A)=\ln A/\ (1-q),$ then 
\begin{equation}
S=\frac{\ln \left( \sum_{\alpha }p_{\alpha }^{q}\right) }{1-q}  
\label{Reny}
\end{equation}%
is the Renyi entropy. If instead we take $f(A)=(A-1)/(1-q),\ S$ corresponds
exactly to the Tsallis entropy (\ref{STsallis}). This indeterminacy is
obvious also from Eq. (\ref{min}), since the constraint term $\lambda \left(
\sum p_{\alpha }^{q}-A\right) $ can also be taken as 
$\lambda ^{\prime }\left[ f\left( \sum p_{\alpha }^{q}\right) -f(A)\right]$
without changing its significance, for any single-valued function $f$. Thus,
it is clear that the exact form of $f$ is not essential in our approach.

We should stress that the thermodynamical properties of the preequilibrium
state discussed above are not the nonextensive ones discussed by
Tsallis~et~al. \cite{Tsallis,TsallisRev}. There, the thermodynamical and
statistical properties for $q$-modified values are studied. In the present
approach, the $q$-modified quantities refer to the subsystem of $q$%
-clusters. When we want to calculate, for example, the internal energy of
the system, the relevant quantity is not $\left\langle E \right\rangle _{q}$
but just the usual total energy $E$. Therefore, the effective equation of
state, that is, the relation between pressure and energy density given by
Eq.~(\ref{P}), will apply immediately as the real physical relation which
can be used, for example, in hydrodynamical calculations. However, the
``entropy'' here is not the real entropy of the whole system but just
reflects the total number of clusters among correlated particles. Thus, the
time evolution of this quantity is not guaranteed to obey the Second Law of
Thermodynamics, that is, $dS/dt\geq0$. Hereafter, in order to distinguish
from the real entropy, we call this $S$ the ``pseudo-entropy'' of the total
system.

On the other hand, note that the variational equation Eq.~(\ref{min}) can
also be obtained as that of maximizing the cluster multiplicity $M$ at a
fixed average energy per cluster. We would have the variational equation, 
\begin{equation}
\delta\left[ M+\beta\left( \left\langle E \right\rangle _{q} -E_{C}\right)
+\nu\left( \sum_{\alpha}p_{\alpha}-1\right) \right] =0 \,,
\end{equation}
where $E_{C}$ is the average energy per cluster, and $\beta$ and $\nu$ are
Lagrange multipliers. This leads to the same equation as Eq.~(\ref{min}).
Since $M$ is the total number of possible ways to form clusters, we can
define the ``entropy'' of the system of clusters by $S_{q}=\ln M,$ and this
corresponds to the Renyi entropy, Eq.~(\ref{Reny}). Therefore, Eq.~(\ref{min}) 
is nothing but the maximization of the entropy defined only for the
restricted subspace corresponding to the degrees of freedom of clusters in
the entire phase space of the system.

In the above discussion, we considered the value of $q$, i.e., the number of
strongly correlated particles as constant for any single-particle state $%
\alpha $. However, in a more general case, they may be different for each
single-particle state and may even depend on the proper occupation numbers $%
p_{\alpha }$. Therefore, for a more general form of the correlation pattern,
the number of cluster states $M_{\alpha }$ should be written as 
\begin{equation}
M_{\alpha }=M_{\alpha }\left[ \left\{ p_{\alpha }\right\} \right] =\frac{%
p_{\alpha }^{F_{\alpha }\left[ \left\{ p_{\alpha }\right\} \right] }}{\Gamma
\left( F_{\alpha }\left[ \left\{ p_{\alpha }\right\} \right] +1\right) }
\end{equation}%
where $F_{\alpha }\left[ \left\{ p_{\alpha }\right\} \right] $ are
functionals of $\left\{ p_{\alpha }\right\} .$ Then, the Eq.~(\ref{min})
should have the form 
\begin{equation}
\delta \left[ \frac{\sum M_{\alpha }\left[ \left\{ p_{\alpha }\right\} %
\right] E_{\alpha }}{\sum M_{\alpha }\left[ \left\{ p_{\alpha }\right\} %
\right] }+\lambda \left( \sum M_{\alpha }\left[ \left\{ p_{\alpha }\right\} %
\right] -M\right) \right. 
\left. +\nu \left( \sum p_{\alpha }-1\right) \right] = 0 \, . 
\label{min_geral}
\end{equation}%
The resulting probabilities $p_{\alpha }$ will not be of the form of Eq.~(%
\ref{p_a}) anymore, yet functions of $V$, $N$, and $M$. This suggests that
the preequilibrium state can be attained for a wide variety of
single-particle spectra and the corresponding pseudo-entropy will be given
as a function of $M$, 
$ S=f\left( M\right) =f\left( \sum_{\alpha }M_{\alpha }\right)$.
This is presumably the most general form of the pseudo-entropy. However, we
should recall that it does not necessarily possess the properties of entropy
required by the usual thermodynamics because the state specified by Eq.(\ref%
{min_geral}) does not correspond to the conventional equilibrium of the
system.

\section{Example}

Discussion in the previous section indicates what kind of dynamical
correlations may lead to Tsallis type occupation probability distribution.
Just to see a concrete example let us consider a very simple toy model
described below.

\begin{enumerate}
\item Initially $N$ particles are distributed as $\left\{
N_{0},N_{1},...,N_{\alpha },...\right\} $ over equally spaced energy levels.
Let us denote the energy of the $i-$th particle as $E_{i}.$

\item Choose randomly a pair of particles, say, $i$ and $j.$

\item Energies of $i$ and $j$ are updated according to one of the following
alternatives:

\begin{enumerate}
\item The new energies are set to the lower one of $E_{i}$ and $E_{j},$ that
is, 
$E_{i},E_{j}\rightarrow E_{i}^{\prime}=E_{j}^{\prime}=Min\left(
E_{i},E_{j}\right)$,
then choose another $k$-th $\left( k\neq i,j\right) $ particle randomly and
attribute 
$E_{k}^{\prime}=E_{k}+Max\left( E_{i},E_{j}\right) -Min(E_{i},E_{j})$
to conserve the total energy.

\item Change the energies as 
$E_{i} \rightarrow E_{i}^{\prime}=E_{i}\pm\Delta E$ and 
$E_{j} \rightarrow E_{j}^{\prime}=E_{j}\mp\Delta E$,
where $\Delta E$ is the level spacing. Here, if one of 
$E_{i}^{\prime},E_{j}^{\prime}$ becomes negative, then 
this step is skipped. That
is, we have to keep always $E_{i}\geq0$.
\end{enumerate}

\item The alternatives ($a$) and ($b$) are chosen randomly, but the ratio of
the average frequency of ($a$) to ($b$) is kept as a constant, $r$.
\end{enumerate}

It is well-known that for $r=0$ in the above model, the ultimate
single-particle spectrum will be the Boltzmann distribution. When $r>0$, the
collisions of type ($a$) are forming a kind of cluster, while the collisions
of type ($b$) may destroy a correlation formed before. Thus, we expect the
number of correlated particles will reach some stationary value.
Furthermore, by construction, the energy of a correlated pair always tends
to diminish. In this way, we may expect that the above system will lead to
the situation described by Eq.~(\ref{min}). In Fig.~2, we show the results
of simulations, for several values of $r$ from $0.00001$ to $0.1$. It is
interesting to note that $r>0$ leads to a non-Boltzmann distribution which
is well approximated by the Tsallis distribution, $p_{a}=C/\left[ 1+\left(
q-1\right) \beta E_{\alpha }\right] ^{1/(1-q)}$, where $\beta
=1/(3-2q)\left\langle E\right\rangle $ and $C=(2-q)/(3-2q)\left\langle
E\right\rangle $ are determined from the normalization condition and
conservation of energy. One parameter fits with respect to $q$ were
performed and the results are indicated by the continuous curves in this
figure. For $r\rightarrow 0$, the spectrum converges to the Boltzmann
distribution $p_{a}\rightarrow e^{-E_{\alpha }/\left\langle E\right\rangle }$
as expected. Note that for a one dimensional case like the present model,
the Tsallis distribution is valid only for $q<3/2$, otherwise the energy
expectation value diverges. For larger values of $r$, the fitted value of $q$
tends to this limiting value, but the distribution begins to deviate
substantially from the Tsallis distribution in the low energy region.

\begin{figure}[htb]
\begin{center}
\includegraphics[width=6.0cm]{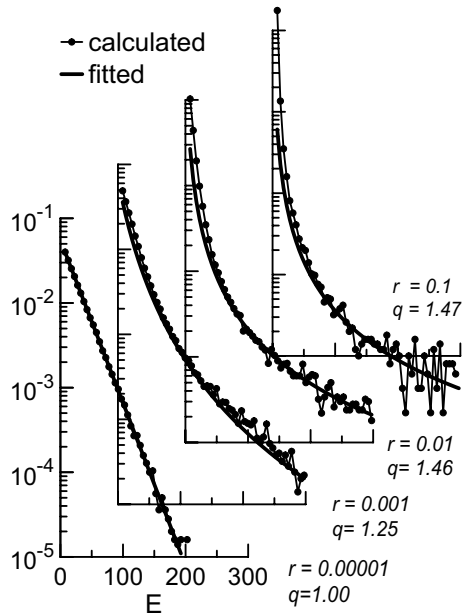}
\caption{Energy spectrum after a large number of collisions per particle,
starting from a distribution peaked at $\left\langle E\right\rangle =22.5$.
The probability of cluster-forming collisions ranges from $r=0.1$ (top) to $%
r=0.00001$ (bottom). Tsallis distributions fitted to the calculated spectra
are also shown, together with the corresponding $q$-values. }
\end{center}
\end{figure}

It is important to note that these distributions are the stationary and
stable ones. We confirm that starting from any different initial conditions,
the system always converges to the same final distribution, uniquely 
determined by the $r$ parameter.
Furthermore, the convergence for $r>0$ is much faster than for $r=0$.
For example, taking $r=0.01$ we find that the distribution converges 10 
times faster than for $r=0$.

In the above toy model, time reversal is violated in type ($a$) collisions. 
However, this is not the crucial factor to obtain the non-Boltzmann
distribution. We have checked this in a more elaborate model which has
time-reversal invariance.

\section{Conclusions and perspectives}

We have shown that a very simple mechanism of dynamical correlations among
particles may lead to a state described by the so-called nonextensive
statistics of Tsallis. It is shown that the nonextensive statistics can be
interpreted as the statistical mechanics for a subspace defined by these
strongly correlated clusters in the entire phase space of the whole system.
Furthermore, the requirement of the minimum energy per cluster or the
maximum \textquotedblleft entropy\textquotedblright\ of the subspace defines
the thermodynamical properties of the whole system too. There, the energy is
defined by the usual sum of single-particle energies multiplied by their
occupation probability, and not by the $q$-weighted average of the energy.
We have also shown a toy model for which such a mechanism actually takes
place.

In this example, the stationary state is attained with a single-particle
spectrum which is well represented by that of Tsallis statistics. These
stationary non-Boltzmann distribution are attained much earlier than the
case of no correlations, hence the Boltzmann equilibrium. The reason for
reaching the stationary distribution much earlier than the real thermal
equilibrium is that the formation of correlated clusters serves as a kind of
catalyzer to distribute the energy among particles more efficiently,
involving more than 2-body collisions. In this aspect, the dynamical
mechanism responsible for the non-Boltzmann distribution is similar to that
of the kinematical approach discussed by Rafelski~et~al. \cite{Rafelski}.
Many examples of systems for which Tsallis statistics have been applied may
also be interpreted in our scheme easily. We also found that this
non-Boltzmann equilibrium may happen not only with a power type but with a
much wider class of single-particle spectra. 

The present work is rather phenomenological and so the conclusion is still
speculative. More theoretical work based on realistic field theoretical
models should be developed. Further investigations in this line are in
progress. This work has been supported by CNPq, FINEP, FAPERJ, and
CAPES/PROBRAL.

\end{document}